# Real-Space Imaging of the Ordered Small Molecule Orientations in Porous Frameworks by Electron Microscopy

**Authors:** Boyuan Shen[1], Xiao Chen[1]*, Dali Cai[1], Hao Xiong[1], Shifeng Jin[2], Xin Liu[3,4], Yu Han[4,5], Fei Wei[1]*

**Affiliations:**
[1]Beijing Key Laboratory of Green Chemical Reaction Engineering and Technology, Department of Chemical Engineering, Tsinghua University, Beijing 100084, China.
[2]Research and Development Center for Functional Crystals, Beijing National Laboratory for Condensed Matter Physics, Institute of Physics, Chinese Academy of Sciences, Beijing 100190, China
[3]School of Chemistry, Dalian University of Technique, Dalian 116024, China
[4]KAUST Catalysis Center, Physical Sciences and Engineering Division, King Abdullah University of Science and Technology, Thuwal 23955-6900, Saudi Arabia.
[5]Advanced Membranes and Porous Materials Center, Physical Sciences and Engineering Division, King Abdullah University of Science and Technology, Thuwal 23955-6900, Saudi Arabia.
*Corresponding authors. Email:
chenx123@tsinghua.edu.cn (X.C.); wf-dce@tsinghua.edu.cn (F.W.)

Abstract
The real-space imaging of small molecules is always challenging under the electron microscopes, but highly demanded for investigating various nanoscale interactions, such as hydrogen bond and van der Waals (vdW) force[1-6]. Especially, identifying the host-guest interactions in porous materials directly at the molecular level will bring a deeper insight into the behaviors of guest molecules during the sorption, catalysis, gas separation and energy storage. In this work, we directly resolved the ordered configurations of p-xylenes (PXs) adsorbed in ZSM-5 frameworks by the scanning transmission electron microscopy (STEM) with the integrated differential phase contrast (iDPC) technique[7-10] to identify the host-guest vdW interactions. Based on these observations, we revealed that the PXs in one straight channel modified the channel geometry with a coherent orientation. And the adjacent straight channels were deformed up to 8.8% along the different directions corresponding to three dominant PX configurations, resulting a negligible overall expansion of ZSM-5 lattices. Then, we could also image the disorder and desorption of PXs in ZSM-5 channels during the in situ heating. This work not only helped us to study the host-guest vdW interactions and the sorption behaviors of PXs in ZSM-5, but also provided an efficient tool for further imaging and studying other single-molecule behaviors under STEMs.

Investigating the nanoscale interactions is one of the most important issues in the fields of nanotechnology. For example, the van der Waals (vdW) interactions between the host framework and the guest molecules are significant in various macroscopic applications, such as the sorption, catalysis, gas separation and energy storage[11-17], but difficult to be studied experimentally at the molecular level. The previous studies based on collective methods, such as the powder X-ray/neutron diffraction (PXR/ND), nuclear magnetic resonance (NMR) and simulations[18-22], can only provide the averaged information of bulk materials, not considering the inevitable differences in individual channels/pores and local structures. The real-space imaging of small molecules helps us studying the molecular interactions at the single-molecule level[1-6], and reveals the binding sites and configurations of molecules as pointers to detect the host-guest interactions. However, the single molecules imaged by the atomic force and scanning tunnel microscopy were mainly on the substrate surfaces. For the molecules adsorbed in frameworks, imaging them under electron microscopes is a feasible pathway, but is still challenging owing to the insufficient resolution and natural electron-beam damage.

Recently, some strategies to image beam-sensitive materials by (scanning) transmission electron microscopes ((S)TEMs) were rapidly developed[23-25]. Among them, the STEM with the integrated differential phase contrast (iDPC) technique[7-10] provide a feasible tool to resolve small molecules in host frameworks. The iDPC-STEM images can be obtained under a low beam current with high resolution and signal-to-noise ratio, and the contrasts of light elements were enhanced compared to other imaging modes. In this work, we achieved the real-space imaging of p-xylene (PX) pointers confined in the ZSM-5 frameworks[26] (Fig. S1) by the iDPC-STEM. The clearly imaged configurations were used to identified the vdW potentials in different channels of ZSM-5. Based on these new observations, the host-guest interactions in zeolite catalysts and the sorption behaviors of aromatics were re-understood at the molecular level.

Fig. 1a exhibits the schematic diagram of our experiments. The specimens we used to adsorb aromatics were the short-b-axis ZSM-5 crystals with the coffin-like shapes and smooth (010) surfaces (Fig. S2). And the PXs were adsorbed in the straight channels of ZSM-5 directly from the liquid phase and used as the pointers to indicate the vdW potentials inside. Then, using a Cs-corrected STEM with iDPC technique, we imaged not only the ZSM-5 framework but also the arrays of light-element PXs in its channels with ultimate resolution. In the set-up of iDPC-STEM (Fig. S3), a segmented annular detector with 4 quadrants is used to detect the electrons and obtain the iDPC-STEM images after taking a 2D integration on the DPC images.

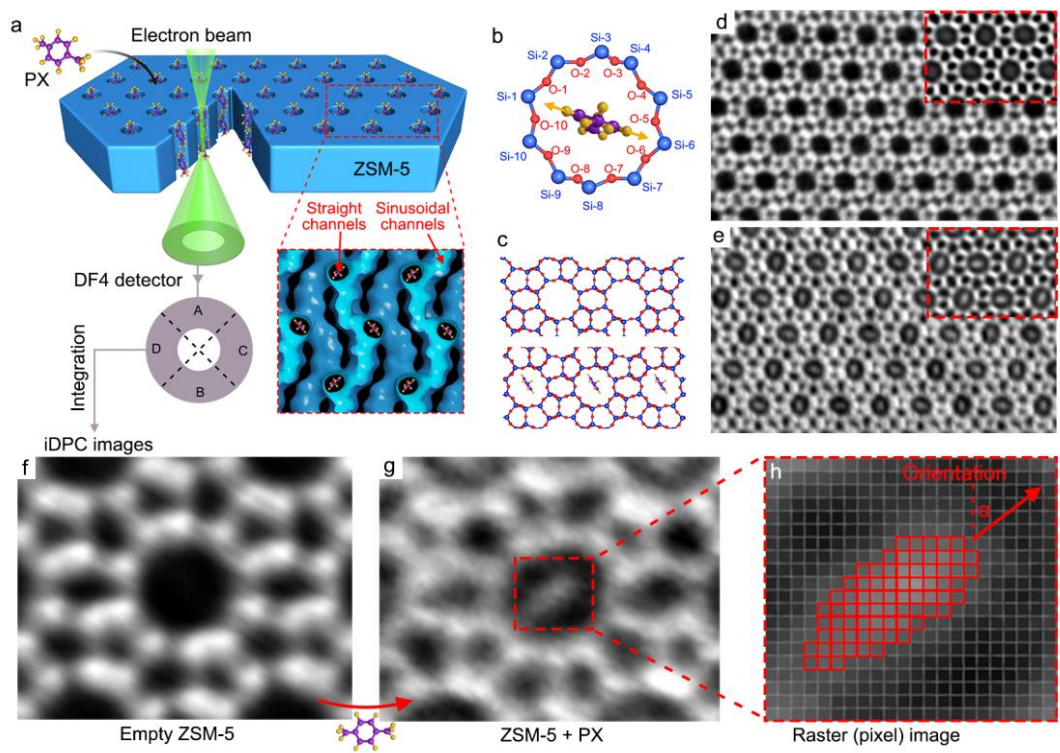

**Fig. 1 | The iDPC-STEM imaging of PX pointers confined in ZSM-5. a**, The schematic diagram of confining and imaging PXs in ZSM-5. **b**, The PX pointer in a numbered $Si_{10}/O_{10}$-ring. **c**, The structural models of the ZSM-5 with and without PXs. **e, f**, The iDPC-STEM images of the empty ZSM-5 and PX-filled ZSM-5 from the [010] projection. The insets show the simulated projected electrostatic potential, consistent with the imaging results. **f, g**, The magnified iDPC-STEM images of the empty and PX-filled ZSM-5 respectively. **h**, The original raster image of the PX spot in **g** showing the intensity of each pixel. The shape and orientation of this PX spot can be defined. (pixel size: 0.2514×0.2514 Å$^2$)

From the [010] projection, the straight channels of ZSM-5 were formed by the elliptical $Si_{10}/O_{10}$ rings numbered in Fig. 1b. And one straight channel (with the size of ~5.6×5.3 Å$^2$) could just contain single array of $C_6$-rings (with the kinetic diameter of ~5.8 Å). In the previous adsorption study, the probable averaged positions of guest molecules in ZSM-5 can be revealed by the mentioned methods[27-32]. The monocyclic aromatics (PXs for example) could be stably confined in the straight channels of ZSM-5 with their long molecular axes nearly parallel to the straight channel axis. And its $C_6$-ring shows an energetically preferred orientation due to the interactions with the frameworks (Fig. 1c and S4, Text S3). Thus, the rotating aromatic pointers can be used to detect the vdW potential field formed by the $Si_{10}/O_{10}$-rings of these channels. And they can be directly imaged using different imaging modes in the STEM (Fig. S5 and S6). Among all these imaging modes, the iDPC-STEM exhibited the highest imaging resolution indicating an information transfer of 1.2 Å to display the shapes and orientations of PX spots more clearly.

Fig. 1d and e compared the iDPC-STEM images of empty and PX-filled ZSM-5 from the [010] projection. It is obvious that, after the PX adsorption, there is a spindle-shaped spot oriented in each straight channel, which is highly consistent with the simulation results in the insets. Fig. 1f and g show the typical single channel with and without PXs respectively. And in the raster image in Fig. 1h, we can outline the shape of PX spot by the intensity isoline at half maximum of intensity peak in the channel (as marked by the red frames, Text S2). The observed sharp PX shape indicates that an array of $C_6$-planes along the straight channel shows a coherent orientation which remained stable within the dwell time of beam scanning (at least 90 ms per channel). Such ordered packing arrangement of PXs in the ZSM-5 has been widely reported during the PX adsorption at low loadings (~four PXs per unit cell)[27] due to the confinement effect of ZSM-5, the interactions of methyl H atoms and the entropy contributions. Then, the orientation of this spot was defined as its long-axis direction and expressed as the acute angle α between this orientation and the a-axis of ZSM-5.

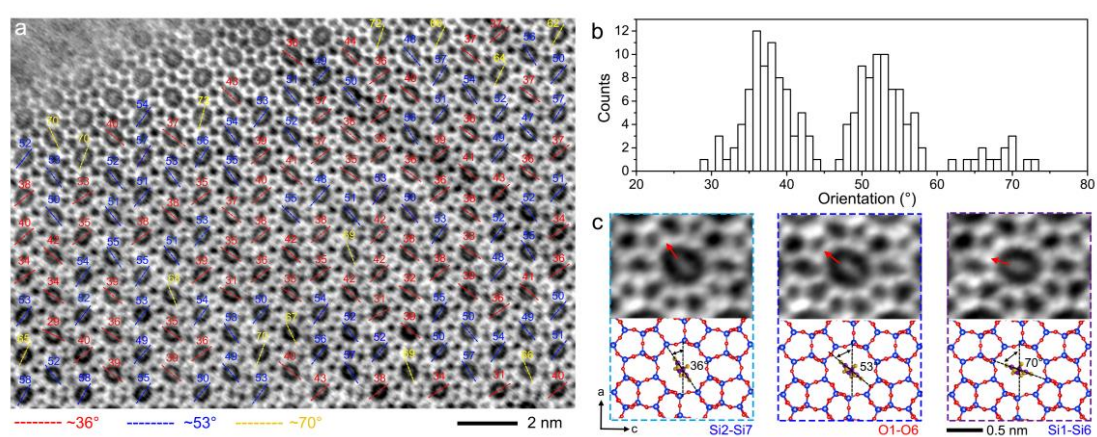

**Fig. 2 | The orientations of PX pointers in ZSM-5. a**, The iDPC-STEM image of PX-filled ZSM-5 showing the orientations of PXs in different straight channels as marked by the colorful lines. **b**, The orientation distribution of PX pointers in the straight channels of ZSM-5 based on the data in **a**. And three separated peaks were observed at around the angles of 36, 53 and 70°. **c**, The magnified iDPC-STEM images of PX spots with three orientations at around 36, 53 and 70° respectively. (pixel size: 0.2514×0.2514 Å$^2$)

The PX orientations visually indicated the interactions between the host ZSM-5 and the guest PX molecules. After counting the orientations of PX spots in the channels shown in Fig. 2a, we found the orientations of PX arrays were varied in different channels and randomly distributed in the STEM image from the [010] projection. The histogram of PX orientations (Fig. 2b) shows a regular distribution with three discrete peaks at about 36, 53 and 70°, which are considered as three dominant stable configurations of PXs in the ZSM-5 channels (as marked by different colors in Fig. 2a). And more statistics with the PX orientations in over 1000 channels (Fig. S7) also confirmed these results. In Fig. 2c, the magnified iDPC-STEM images illustrated the PX spots with three dominant orientations compared to the probable structural models. With these orientations, the vertical $C_6$-rings pointed to the different opposite Si and O atoms on $Si_{10}/O_{10}$ rings (Si2-Si7, O1-O6 and Si1-Si7 respectively). Meanwhile, the sharp shapes of all these spots indicated that the whole array of PXs (at least most of them) changed their orientations, which may also reveal the diverse vdW interactions in different channels.

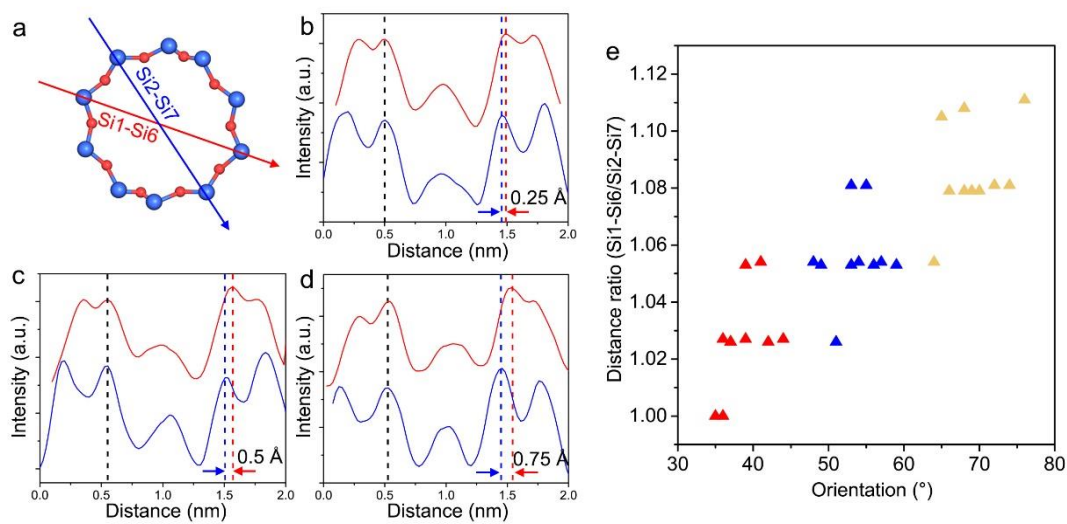

**Fig. 3 | The changes of channel geometry corresponding to the PX orientations. a**, The profile directions to measure the Si1-Si6 and Si2-Si7 distances. **b-d**, The intensity profiles of three iDPC-STEM images in Fig. 2c with different PX orientations respectively. The differences between the projected Si1-Si6 and Si2-Si7 distances indicate the different changes of channel geometry. **e**, The correlation between the ratios of projected Si1-Si6 to Si2-Si7 distances and the PX orientations.

After atomically analyzing the ZSM-5 framework, we attributed such distributions of orientations and vdW interactions to the different modification of channel geometry. It is well known that the adsorbed molecules caused the reversible structural changes of ZSM-5 framework at very low loadings (less than one PX per unit cell), confirmed by the $^{29}$Si MAS-NMR and PXRD results[27-30]. And we have also provided the real-space proof by the profile analysis of STEM images[10]. Here we further found that the adjacent channels exhibit the different deformations to induce the different orientations of inside PXs. We measured the projected Si1-Si6 and Si2-Si7 distances from two intensity profiles marked in Fig. 3a to estimate the changes of channel geometry. The intensity profiles and measured results of three channels in Fig. 2c were given in Fig. 3b-d. The difference between the Si1-Si6 and Si2-Si7 distances described the geometry of $Si_{10}$-rings with different PX orientations. Fig. 3e gives the summarized results of the ratios of Si1-Si6 to Si2-Si7 distances related to the PX orientations (more data in Fig. S9-11), indicating that these $Si_{10}$-rings were obviously lengthened in the PX-point directions (compared to the results of empty ZMS-5 in Fig. S8). Such channel deformation can be as large as 8.8%, while the overall expansion of the ZSM-5 lattices is negligible.

Thus, during the adsorption of PXs with a larger kinetic diameter than the channel size, the deformation of straight channel provided the proper space for each channel to stably confine the array of PXs with a coherent orientation. And these $Si_{10}$-rings of straight channels were deformed along different directions, which is corresponding to a natural distribution of PX orientations with three dominant peaks in Fig. 2b, to avoid the overall expansion of ZSM-5 lattices and maintain the structural stability of frameworks. The largest deformation of 8.8% corresponding to the PX orientation of near 70° obviously existed with the least possibility. More importantly, the relation between the molecular configuration and the channel geometry can only be revealed by the real-space imaging, which visually explained the strong host-guest interaction and adsorption ability of PXs in ZSM-5 channels.

Then, we investigated the desorption behavior of PXs by the in situ heating experiment in STEM. Fig. 4a-f and Fig. S12 show the iDPC-STEM images of PX-filled ZSM-5 at increasing temperatures from 50 to 300 °C. And the profile analysis along the c-axis of ZSM-5 displayed the peak intensity in each profile valley to indicate the contrast of PX spot (marked by the colored bands in given profiles). As we observed in these profiles, the contrasts (peak intensities) of PX arrays barely changed before heating to 250 °C, which obviously decreased in the profiles at higher temperatures. And the appearance of empty channels at over 250 °C indicated that the PXs gradually escaped from the channels to the vapor phase, which agrees with the thermal gravimetric analysis (TGA) results of PX desorption in ZSM-5 (Fig. S13).

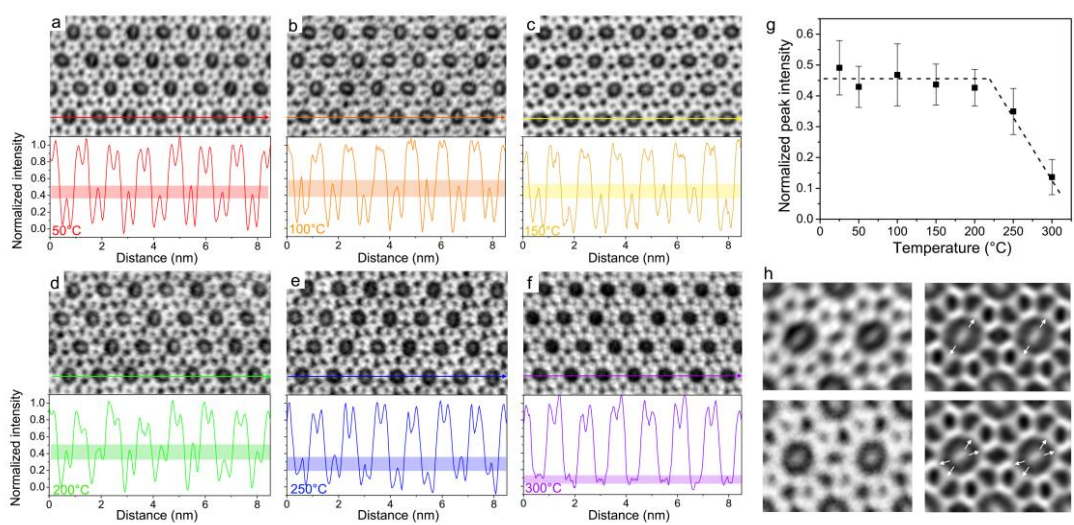

**Fig. 4 | The thermal-induced PX desorption during the *in situ* heating process. a-f**, The iDPC-STEM images and the corresponding normalized profile analysis of PX-filled ZSM-5 obtained at different temperatures. **g**, The average peak intensities in the profiles at different temperatures. The obvious decrease from 200 to 300 °C indicating the desorption of PXs from the ZSM-5 channels. **h**, The magnified iDPC-STEM images showing the transition from the single-oriented PX spots with the sharp elliptical shapes to the multi-oriented ones with the round shapes after heating up to 200 °C.

When checking the imaged PX arrays in details (Fig. 3h), more and more PX spots no longer have a sharp shape and single orientation at over 200 °C. Just before the PX desorption, the ordered packing of PX turn into the multi-oriented state with a nearly round shape due to their thermal-induced rotation and oscillation near the equilibrium positions (perhaps together with the faster changes of channel geometry) during the dwell time of probe scanning. Thus, the PX desorption occurred in a high correlation with the destruction of configuration stability by the increased thermal energy. These observations provided us a new understanding on the dynamic desorption behaviors of guests at the molecular level.

In conclusion, the iDPC technique helped us imaging the single-molecule-sized aromatics and revealing the interactions between the aromatics and ZSM-5 frameworks directly under the STEM. After the PX adsorption, the PXs induced different structural changes in adjacent channels to avoid the overall expansion of ZSM-5 lattices. And the PXs in these channels could act as pointers to sensitively detect the slight changes of vdW potentials corresponding to the channel geometry and show the energetically preferred orientations. These results allow us to understand the adsorption/desorption of PXs in ZSM-5 frameworks from the perspectives of molecular configuration, which can only be studied by the real-space method due to the aperiodicity of structure. Moreover, other small organic molecules, such as homologues of benzene, can also be confined in the ZSM-5 and other size-matching porous zeolite frameworks. Thus, this work provides a general method to study their interactions in a series of organic/inorganic systems by the real-space evidences, which has attracted wide research interests in the sorption and catalysis processes, however, difficult to be investigated at the atomic scale. And such progress on imaging technique will further promote the study on the physical and chemical properties of these guest molecules, bringing new insights on diverse single-molecule behaviors.

**Materials and Methods**

Synthesis of short-b-axis ZSM-5 crystals

The short-b-axis ZSM-5 crystals we used were synthesized via a hydrothermal method[31,32]. The reactants, including tetrapropylammonium hydroxide (TPAOH, 25% in water, 13.1 g), tetraethylorthosilicate (TEOS, 11.2 g), iso-propanol (IPA, 0.1 g) urea (2.0 g), Al(NO$_3$)$_3$·9H$_2$O (0.3g) and NaOH (0.1 g), were added in 18.4 g H$_2$O and fully mixed at ambient conditions over 2 h. For the further crystallization, the mixed solution was heated to 180 °C with a rate of 15 °C/h in an autoclave. And then, after another 50 h keeping at 180 °C, the autoclave was put into cold water to quench the crystallization. After washing away impurities by deionized water, the templates were removed by calcining at 500 °C for over 5 h. Then, the as-prepared Na-ZSM-5 turned into the used empty ZSM-5 via another two process, including ion exchange in NH$_4$NO$_3$ solution and H$_2$-protected calcination.

Filling the ZSM-5 framework with PX molecules

PX molecules were absorbed into ZSM-5 straight channels directly from the liquid phase. Some ZSM-5 powder and pure PX liquid were added into a 1-mL centrifuge tube. The ZSM-5 crystals were dispersed in PX liquid using ultrasound for 30 min. The centrifuge tube was stored for a period of time to fully diffuse the PX molecules into the ZSM-5 frameworks. And before the STEM observations, the specimens were dispersed using ultrasound for another 30 min to image the separated ZSM-5 crystals.

The iDPC-STEM imaging

All STEM observations (including iDPC, BF and ADF) were performed using a FEI

Titan Cubed Themis G2 300 operated at 300 kV with convergence semi-angle of 10 mrad. The microscope was equipped with a DCOR+ spherical aberration corrector for the electron probe which was aligned before every experiment using a standard gold sample. The following aberration coefficients were measured as: A1=1.41 nm; A2= 11.5 nm; B2=22.2 nm; C3=2.05 µm; A3=525 nm; S3=177 nm; A4=8.81 µm, D4=2.39 µm, B4=13.2 µm, C5=-3.95mm, A5=295 µm, S5=111µm, and R5=102 µm assuring a 60 pm resolution under normal circumstances. The beam current employed for iDPC–STEM is lower than 0.1 pA (corresponding to an electron dose lower than 40 e$^-$/Å$^2$). The collection angle of BF-STEM is 0~3 mrad, and the collection angles of ADF and iDPC-STEM are 4~20 mrad. During the heating process, the specimens in *in situ* holder (Protochips, Fusion 350) were heated to different target temperatures with an ultrahigh heating rate of 1000 °C/ms. And then, the projected potentials were simulated by the QSTEM software based on the multislice method. The parameters used in simulation were consistent with those in the experiments. The point spread function width was set to 1.5 Å.

low-loaded form. *J. Am. Chem. Soc.* **127**, 7543-7558 (2005).

# Supporting Information

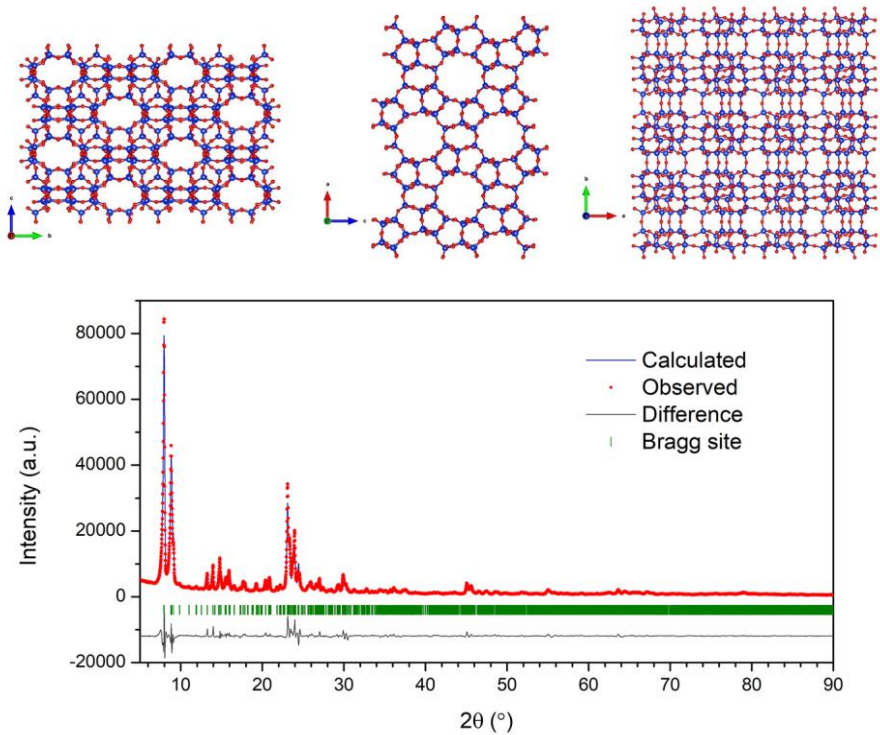

Figure S1 | The structural model of empty ZSM-5 from different projections obtained by the PXRD Rietveld refinement results. The observed (red points), calculated (blue line) results and their difference (black line) were shown in the figure below.

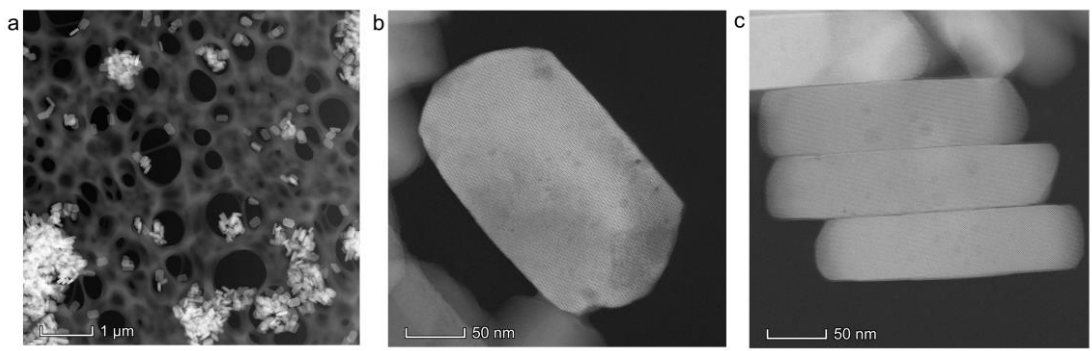

Figure S2 | The ADF-STEM images of short-b-axis ZSM-5 crystals with the coffin-like shapes.

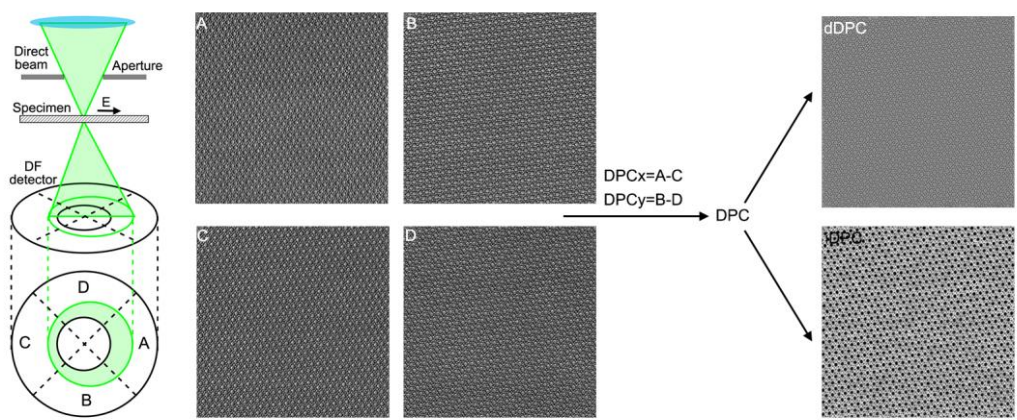

Figure S3 | The schematic set-up for iDPC-STEM imaging and 4 phase contrast images (A-D) detected by 4 quadrants of DF detector respectively. The dDPC and iDPC images were obtained by the 2D differential and integration of DPC image.

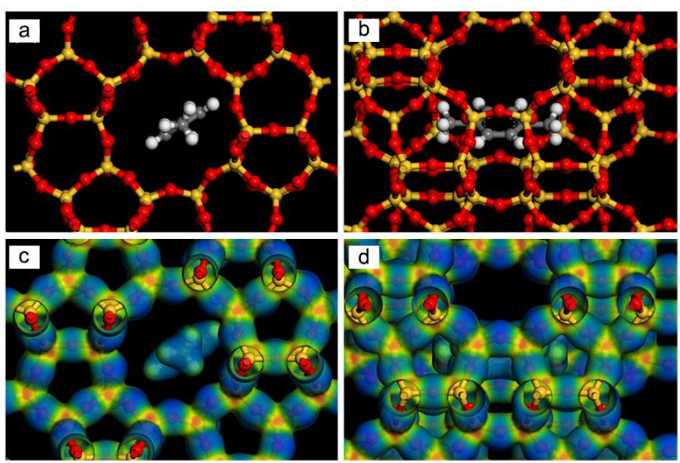

Figure S4 | The most probable configuration of PX molecule in ZSM-5 framework viewed from [010] (**a**) and [100] (**b**) directions. The electrostatic potential mapped onto isosurface of electron density viewed from [010] (**c**) and [100] (**d**) directions. (isovalue is 0.20 a.u. and the potential range is -0.02~0.64 a.u.)

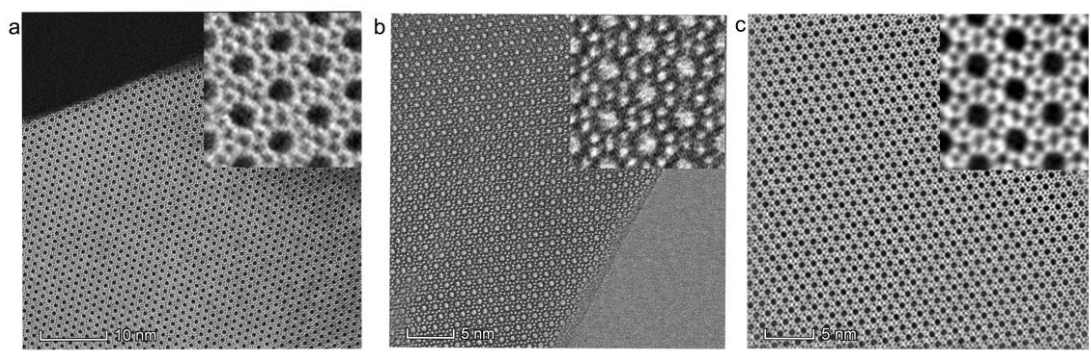

Figure S5 | The ADF (**a**), BF (**b**) and iDPC (**c**) STEM images of the empty ZSM-5.

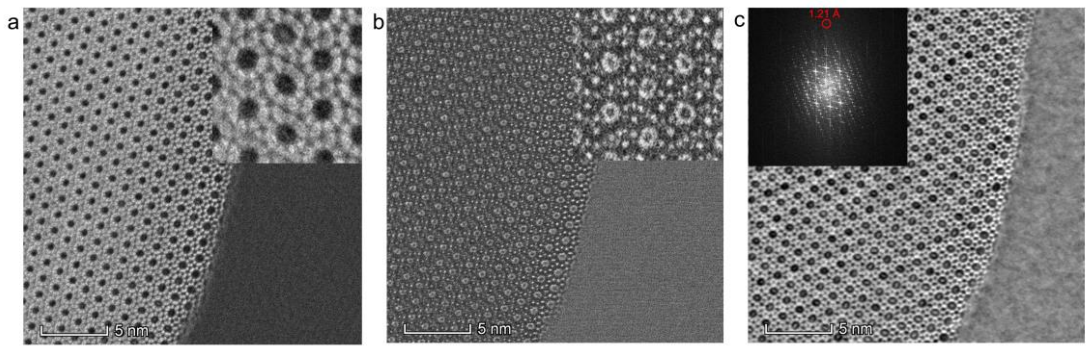

Figure S6 | The ADF (**a**), BF (**b**) and iDPC (**c**) STEM images of the PX-filled ZSM-5 framework at the same place. The iDPC image gives an information transfer of 1.21 Å and shows the clear orientations of PX spots.

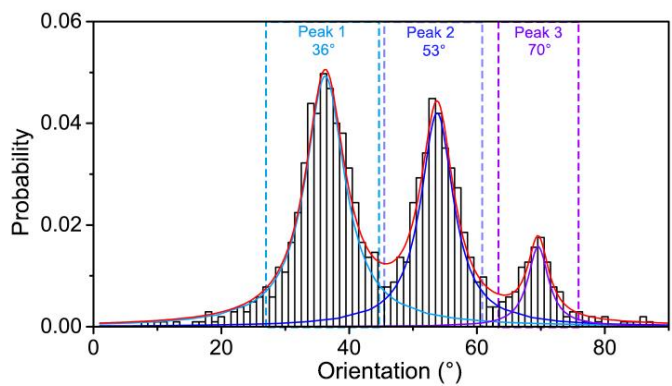

Figure S7 | The orientation distribution of PX spots in the straight channels of ZSM-5 obtained by counting over 1000 PX spots. Three separated peaks were observed at the orientation angles of 36, 53 and 70°, consistent with the results in Fig. 2b.

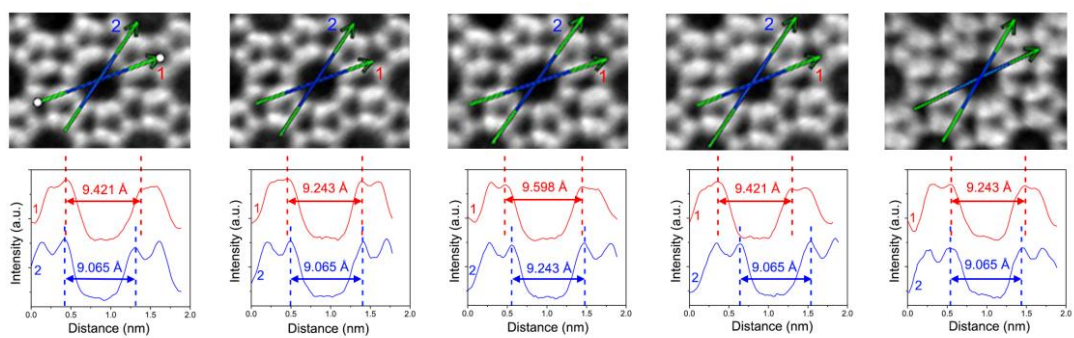

Figure S8 | The PXRD Rietveld refinement of the ZSM-5 after PX filling, including the experiment (red cross points), simulation (blue line), background (orange line) and difference (black line) profiles. The **inset** shows the change of ZSM-5 structures, where the marked distances were measured from the 2D projections from [010] direction.

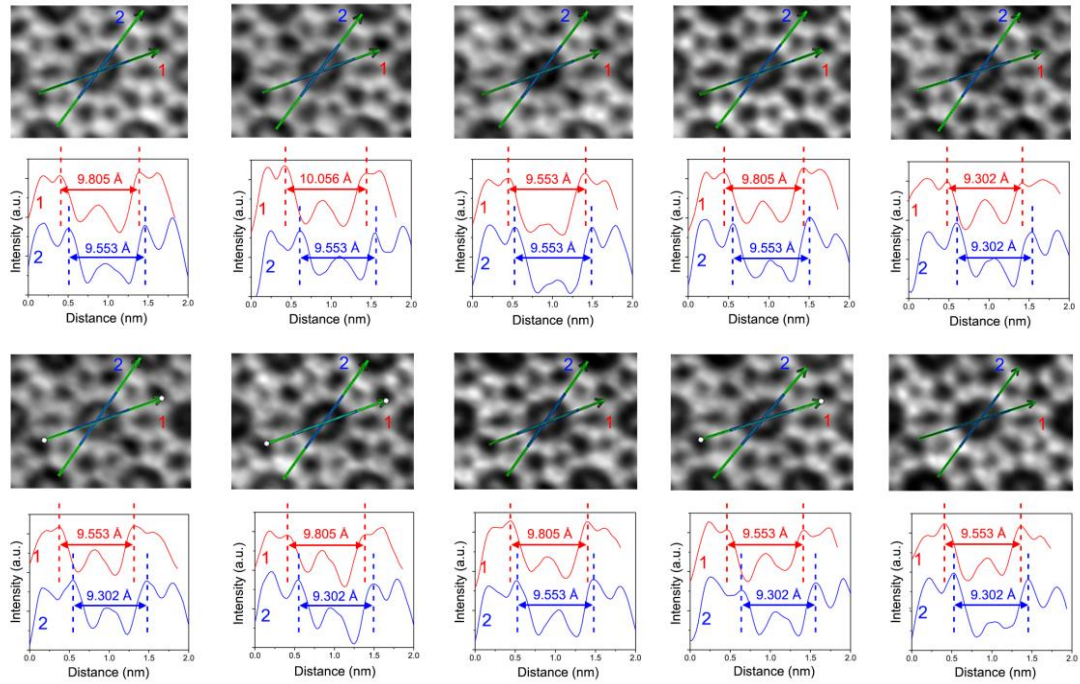

Figure S9 | (**a**) The iDPC-STEM images of the empty and PX-filling ZSM-5. (**b**) The profile analysis of the empty (red) and PX-filling (blue) ZSM-5 along the given profile direction marked in **a**. (**c**) The measured channel dimensions (the distances of two peaks on both sides of the intensity valleys in **b**) indicating the probable change of channel geometry induced by the adsorbed PX.

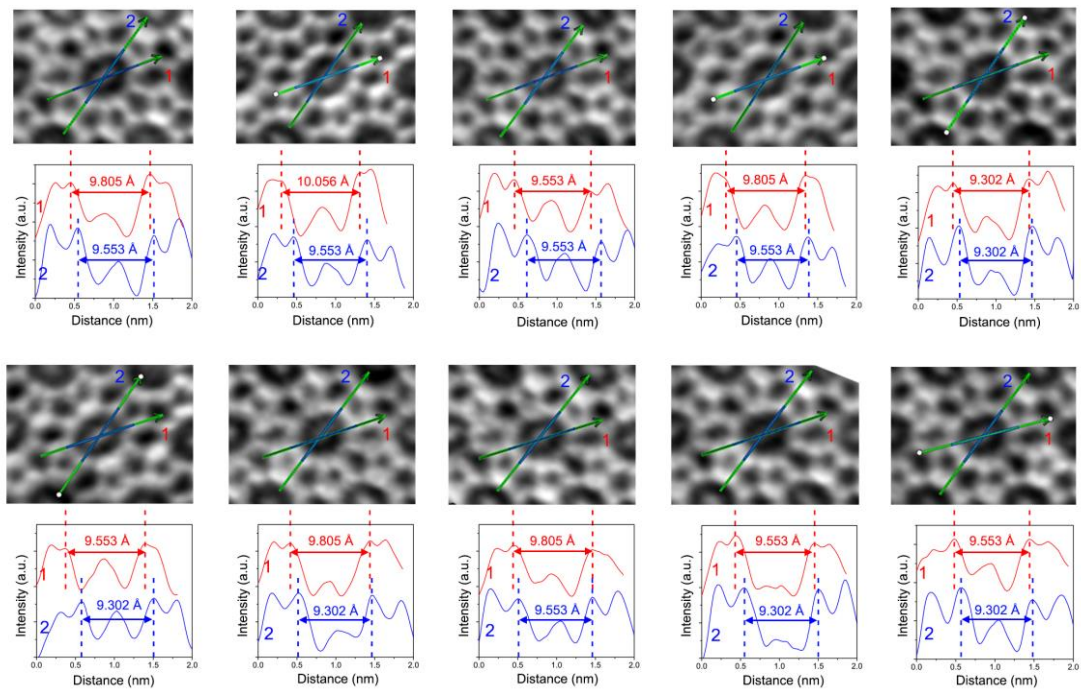

Figure S10 | (**a-c**) Measuring the interatomic distances (numbers of pixels) between Si-1/Si-6 and Si-2/Si-7 from the iDPC-STEM images in Fig. 2g. The average numbers of pixels between Si-1/Si-6 and Si-2/Si-7 of the empty channels are measured as 36 and

35 respectively. (**d**) The ratios of Si-1/Si-6 to Si-2/Si-7 distances at three orientations.

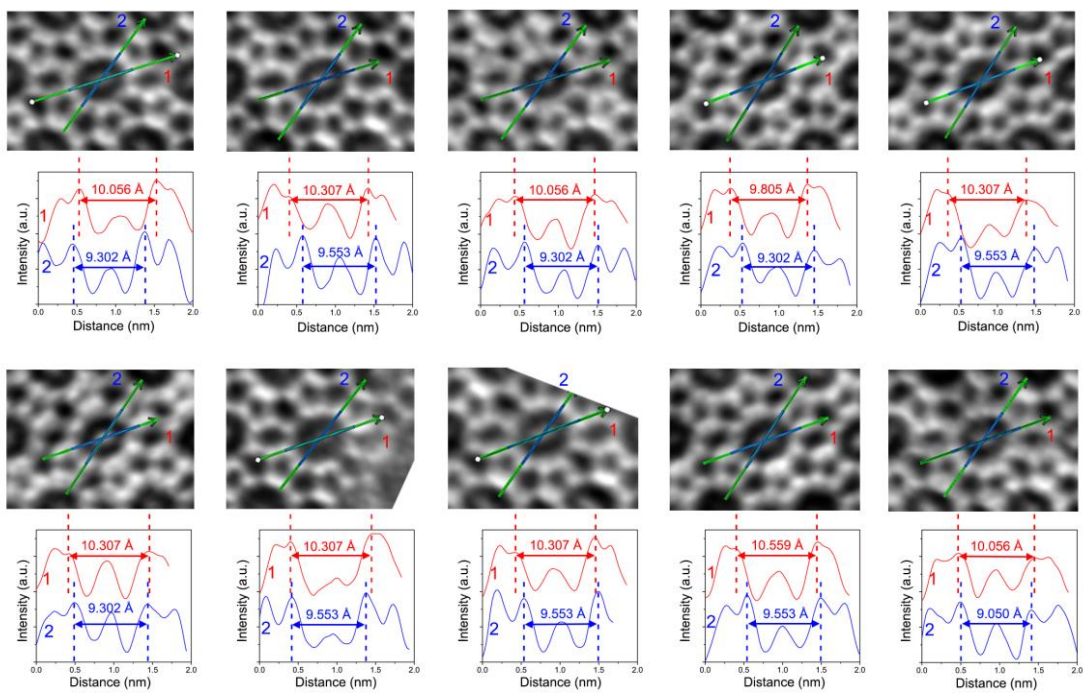

Figure S11 | The iDPC-STEM images of ZSM-5 frameworks filled by PX molecules at different temperatures during the *in situ* heating process.

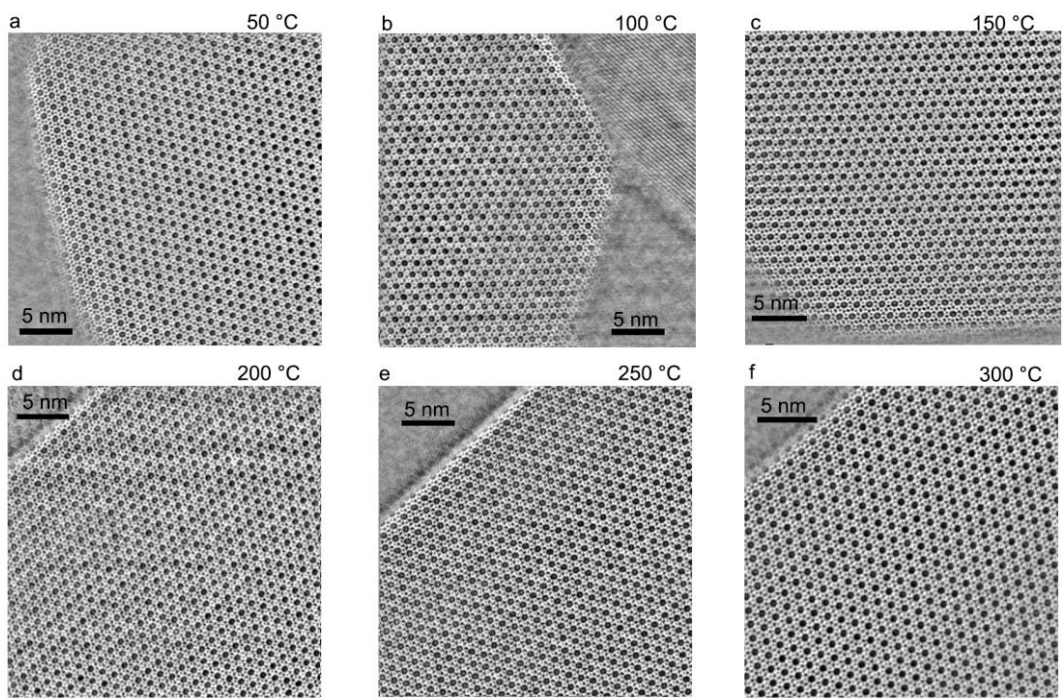

Figure S12 | The iDPC-STEM images of ZSM-5 frameworks filled by PXs at different temperatures during the *in situ* heating process.

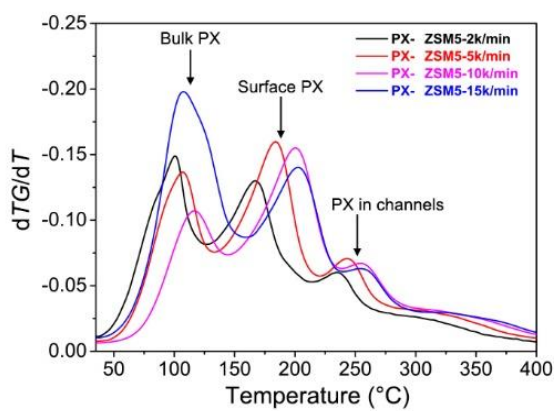

Figure S13 | The thermal gravimetric analysis (TGA) results of PX desorption from the ZSM-5 with different heating rates. Three peaks indicate the phase transition of the bulk PX, the surface PX and the PX confined in ZSM-5 channels respectively.

Text S1
The structural change of ZSM-5 framework
ZSM-5 is a typical MFI-type zeolite material exhibiting an excellent shape selectivity in aromatization reactions due to the unique cross-linked nano-sized channels including the straight channels along b-axis and the sinusoidal channels along a-axis. In this work, we used the PXRD Rietveld refinement to reveal the detailed structure of empty ZSM-5 framework (Fig. S1). First, the PXRD data were collected at room temperature on a PANalytical diffractometer (X'Pert PRO) with Cu $K\alpha$ radiation (40 kV, 40 mA) and a graphite monochromator in a reflection mode ($2\theta$ from 5° to 90°, step=0.017°, scan speed=5 °s/step). Then, the Rietveld refinement was performed using the Reflex module implemented in Materials Studio suite. The previously reported crystal structure model of ZSM-5 (PDF Card - 01-070-4743) was adopted as the starting model for refinement of the empty ZSM-5. The complex structure of ZSM-5 called for additional constrains on Si-O tetrahedron during the Rietveld refinements, and a small fraction of energetic contribution $R_e$ is mixed with the $R_{wp}$ to accounts for the correct solution that with both minimum energy and matching of experimental and simulated patterns. The Universal force field was used during the refinement to give the energetic contribution, and a mixing factor ($R_e/R_{wp}$) of 0.5% is found sufficient to constrain the geometry of Si-O tetrahedron. The refinements readily converged and produced a satisfying fit to the PXRD pattern, with $R_p$ = 7.51% and $R_{wp}$ = 9.87%. Fig. S1 shows the refined structural model of ZSM-5 framework with Si and O atoms viewed from different directions. The as-synthesized ZSM-5 samples showed a high crystallization degree and a uniform coffin-like shape with the atomic-smooth (010) surface and short b-axis (30-50 nm) as shown in Fig. S2. And Fig. S4 gives the high-resolution STEM images showing the clear $Si_{10}$-rings of straight channels viewed from the [010] projection. Such short-b-axis ZSM-5 crystals provided clear templates to investigate the behaviors of absorbed aromatics directly under the STEM at the molecular level.

It has been widely reported that the sorbates will induce reversible structural change of ZSM-5 framework. And in this work, we firstly observed such changes in the real space with the help of iDPC-STEM. Moreover, we also confirmed that the structural changes of different straight channels were different due to the constraints of bond lengths and angles in the framework, which resulted in the different dominant orientations in Fig. 2. As shown in Fig. 3, we choose two profile directions to identify the distances between the opposite Si atoms of $Si_{10}$-rings. And we found that, for the channels with the PXs in three dominant orientations, the ratios of Si1-Si6 to Si2-Si7 distances were different and the correlation between these ratios and orientations were revealed in Fig. 3e. For each dominant orientation, we choose ten channels with very clear $Si_{10}$-rings to obtain the statistics in Fig. 3e and the original data were shown in Fig. S9-11. And compared to the results of empty ZSM-5 channels shown in Fig. S8. It is obvious that, after the PX adsorption, the channels were deformed and lengthened in the directions which the PXs pointed to. These results can be explained by the natural constraints of the lengths and angles of Si-O bonds which connected the adjacent channels. And such structural changes help us better understanding the measured orientation distribution given in Fig. 2b and S7.

Text S2

Using iDPC technique in STEM, we achieved the real-space imaging of PX orientation. In the iDPC-STEM images of PX-filled ZSM-5, there is a spindle-shaped spot in each straight channel with an enough contrast compared to those of the atoms in framework. These spots represented the single-molecule-sized PX species confined in the ZSM-5 channels (arrays of PXs). And we also used the QSTEM software to simulate the [010] projected potentials of PX-filled ZSM-5 specimens (the insets of Fig. 1d and e), which confirmed that the inside PXs could be imaged with proper contrast. Meanwhile, the ADF and BF STEM images also confirm the real-space imaging of PXs as shown in Fig. S6. In the profile analysis of iDPC-STEM image (Fig. 3 and S9-11), the small peak in the intensity valley of each channel indicates the superimposed contrast of PXs along the whole channel. These spots could be treated as near ellipses approximately based on their shapes as shown in Fig. 1h. The spot orientations are defined as the angle between the long axes of ellipses and the a-axis of ZSM-5. Especially, for the spots in the iDPC-STEM images, we used the intensity (grayscale) isoline at half maximum of intensity peak to outline the spot shape. For example, in Fig. 1h, the maximum grayscale of PX spot is 64% (8-bit), thus, the isoline value at half maximum of intensity peak is 64+(100-64)/2 = 82%. Then, we marked the pixels with grayscales less than 82% using the red frames in Fig. 1h. Based on this region, we can outline the shapes of PX spots to identify their orientations. Such method can reduce the influence of the low-intensity signals (noises) and the loss of imaging quality.

The high contrast and sharp shape of spot indicate that an array of PX molecules along the straight channel owns a coherent orientation and shows a superimposed contrast. This orientation is kept stable and unchanged (will not switch between three quantized orientations) during the probe scanning. The dwell time of scanning probe at each pixel is 4 us. And each channel (with the size of ~5.6×5.3 Å$^2$) contains about 22×21 pixels

(the typical pixel size in Fig. 2 and 3 is 25.14×25.14 pm$^2$). Since the whole iDPC image contained 1024×1024 pixels and the probe scanned line by line on the specimens, the dwell time at each channel was calculated as at least 22×1024×0.004 ≈ 90 ms. Thus, within at least 90 ms, the orientation information in single channel was obtained and a stable orientation was detected. Moreover, at higher temperatures (over 200 °C), we observed the multi-oriented spots with nearly round shapes in the channels. And we believe that such transition probably resulted from the thermal-induced rotation and oscillation of PXs (together with the change of channel geometry), while the PXs were activated by heating and the orientations were not as stable as those at room temperature during the dwell time of 90 ms.

Text S3

The first-principles-based calculation was performed using Perdew-Burke-Ernzerhof functional with dispersive corrections as implemented in DMol$^3$. The orthorhombic unit cell of bulk ZSM-5 containing 96 Si atoms and 192 O atoms was used to represent the lattice. The electronic states of Si, C, H and O were treated with DSPP pseudopotentials and DNP basis sets. For all calculations, the real-space global orbital cutoff radius was set to 4.6 Å and the energy convergence criterion was 2×10$^{-4}$ eV. After convergence tests, a 2×2×4 *k*-point grid was used for geometry optimization and electronic structure analysis. With the above setup, the bulk lattice dimension of ZSM-5 was calculated as 20.12×19.90×13.40 Å$^3$ in reasonable agreement with previous theoretical results and the experimental values. The adsorption energy ($E_{ad}$) of guest molecules inside ZSM-5 was calculated as the energy difference between ZSM-5 lattice with PX molecules, the free standing PX molecules in gas phase and the empty ZSM-5 lattice, following the equation:

$$E_{ad} = E_{PX+ZSM-5} - (E_{ZSM-5} + E_{PX}).$$

Among the 51 configurations, including PX absorbed in the straight or sinusoidal channels and their intersections, the PX adsorption within the channel intersection is the most probable. As for the most probable configuration shown in Fig. S7, the $C_{(C6)}$-$C_{(CH3)}$ is parallel to the b-axis of ZSM-5 and the calculated $E_{ad}$ is -104.80 kJ/mol. PX molecules are stabilized by the dispersive interaction formed with the ZSM-5 framework. Structure analysis showed that not only the atoms on the $C_6$-ring but also those in the methyl group are involved in the dispersive interaction. The adsorption of PX is sensitive to the orientation of molecules. If the $C_{(C6)}$-$C_{(CH3)}$ is parallel to the a-axis of ZSM-5, the $E_{ad}$ would increase by at least 11 kJ/mol evenwhen the $C_6$-ring of PX still lying in the intersection. Apart from the intersection, PX adsorption within the straight channel is found less plausible by ~5 kJ/mol, while its adsorption within the sinusoidal channel is found less plausible by at least 20 kJ/mol. In this sense, the possibility for PX residing within the straight channel and at the intersection will be at least 10$^3$ higher than at other positions.